# A Vigorous Explorer Program

*An Activities/Progam White Paper submitted to the Astro2010*
*NAS/NRC Decadal Review of Astronomy and Astrophysics*

Author: Martin Elvis,
Smithsonian Astrophysical Observatory (SAO).
*Phone: (617) 495-7442; email:* elvis@cfa.harvard.edu

Co-Authors:

Matthew Beasley (U. CO, Boulder),
Roger Brissenden (SAO),
Supriya Chakrabarti (Boston U.),
Michael Cherry (Louisiana State U.),
Mark Devlin (U. Penn),
Jerry Edelstein (UC Berkeley),
Peter Eisenhardt (JPL/Caltech),
Paul Feldman (Johns Hopkins U.),
Holland Ford (Johns Hopkins U),
Neil Gehrels (NASA/GSFC),
Leon Golub (SAO),
Herman Marshall (MIT),
Christopher Martin (Caltech),
John Mather (NASA/GSFC),
Stephan McCandliss (Johns Hopkins U.),
Mark McConnell (U. New Hampshire),
Jonathan McDowell (SAO),
David Meier (JPL/Caltech),
Robyn Millan (Dartmouth),
John Mitchell (NASA/GSFC),
Warren Moos (Johns Hopkins U.),
Steven S. Murray (SAO),
John Nousek (Penn State U),
William Oegerle (NASA/GSFC),
Brian Ramsey (NASA/MSFC),

James Green (U.CO, Boulder),
Jonathan Grindlay (Harvard),
Philip Kaaret (U.Iowa),
Mary Elizabeth Kaiser (Johns Hopkins U.)
Lisa Kaltenegger (Harvard),
Justin Kasper (SAO),
Julian Krolik (JHU),
Jeffrey W. Kruk (Johns Hopkins U.),
David Latham (SAO),
John MacKenty (STScI),
Amanda Mainzer (JPL/Caltech),
George Ricker (MIT),
Stephen Rinehart (NASA/GSFC),
Suzanne Romaine (SAO),
Paul Scowen (Arizona State U.),
Eric Silver (SAO),
George Sonneborn (NASA/GSFC),
Daniel Stern (JPL/Caltech),
Mark Swain (JPL/Caltech),
Jean Swank (NASA/GSFC),
Wesley Traub (JPL/Caltech),
Martin Weisskopf (NASA/MSFC),
Michael Werner (JPL/Caltech),
Edward Wright (UCLA)

# Executive Summary

Explorers have made breakthroughs in many fields of astrophysics. The early Explorer program included UHURU, and the restructured, post-1988, program included COBE. The science from both these missions contributed to three of the Nobel Prizes – Giacconi (2002), Mather, and Smoot (2006) - awarded for NASA-based science. Explorers now operating have marked the definitive beginning of precision cosmology, have discovered that short gamma-ray bursts are caused by compact star mergers and have measured metallicity to redshifts $z>6$. NASA Explorers do cutting-edge science that simply cannot be done by facility-class instruments.

The goal of the NASA Explorer program is to provide a rapid response to changing science and technology in order to enable cutting-edge science at moderate cost. In addition, Explorers enable innovation, and engage & train scientists, managers and engineers, adding human capital to NASA and the nation. The astrophysics Explorer launch rate actually being achieved is 1 per 3 years. Explorer budget projections are currently in the $150M/year range for the next five years.

We believe that to enable the program goals, a newly Vigorous Explorer Program should be created, within which Astrophysics Explorers:

1. Reach the long-stated goal of annual astrophysics launches; realization of this goal would have a stongly positive and measurable impact on science and the community;

2. Find additional launch options for Explorers and actively encourage cost savings in launchers and spacecraft, such as new commercial vehicles and innovative partnerships.

3. Mitigate risk via stronger technical development and sub-orbital programs, and through longer and more thorough Phase A programs, potentially reducing the need for a 30% contingency;

4. Strive hard to protect the funding for missions that have reached Phase B, to prevent significant launch slips and cancellations, with a goal of 4 to 5 years from Phase B to launch;

5. Review the project management procedures and requirements to seek cost reductions, including the risk management strategy and the review and reporting process;

6. Review and possibly modify the cost caps for all Explorer classes to optimize scientific returns per dollar;

7. Otherwise follow the recommendations of the 2006 PILMSS NAS/NRC report (Appendix A).

# INTRODUCTION

The 2001 decadal review "Astronomy and Astrophysics in the New Millenium" endorsed the continuation of ***"a vigorous Explorer Program"***. We strongly agree, and go further, by calling for the realization of the long stated goal of an annual astrophysics Explorer launch rate. The 2001 decadal review did not provide a detailed exposition of its endorsement of "a vigorous Explorer Program", and contained only brief statements endorsing a small/medium/large balance of programs. Again, we go further, also calling for a dynamic investigation of ways in which the Explorer program can be made even more effective, such as an adoption of the widest possible range of options to which PI's can *propose*.

The Explorer program goal is to provide rapid response to changing science and technology to enable cutting-edge science at moderate cost. This goal can be broken down into four elements[1]:

> **A.** ***Explorers fill <u>critical science gaps</u> in areas that are not addressed by strategic missions,***
>
> **B.** ***they support the <u>rapid implementation</u> of attacks on very focused topics, and***
>
> **C.** ***they provide for <u>innovation</u> and the use of new approaches that are difficult to incorporate into the long planning cycles needed to get a mission into the strategic mission queues… .***
>
> **D.** ***The Explorers also provide a particularly substantial means to <u>engage and train</u> science and engineering students in the full life cycle of space research projects.***

We detail below both the strengths of the Explorer program, and the challenges that this program faces. We have structured this White Paper around these four 'vital contributions' in order to illustrate the current situation. In this section we address the first, 'Science Gaps'. The 'Technical Overview' section addresses the other three. We draw heavily on the 2006 NAS/NRC report on *"Principal Investigator Led Missions in the Space Sciences"* (PILMSS[2]), which examined many of these challenges.

---

[1] NRC, 2004, *Solar and Space Physics and Its Role in Space Exploration*, Washington, D.C.: The National Academies Press, p. 20. **Bullet letters added.**

[2] http://www.nap.edu/catalog.php?record_id=10949#toc

# KEY SCIENCE GOALS

## The Explorer Program - Overview

The Explorer Program is NASA's oldest flight program dedicated to science investigations. Explorer 1 was launched January 31, 1958. *Swift*, the most recent (launched successfully on November 20, 2004), is the 84th[3] in the series. The 2001 NRC report *"Astronomy and Astrophysics in the New Millennium"* finds that "*the Explorer program is very successful and has elicited many highly innovative, cost-effective proposals for small missions from the community*."

In 1988 the Explorer Program was reconstituted as a competed, PI-led, program, following the PI mode established for the Discovery Program. The introduction of cost caps on the RXTE and ACE missions was part of this transition. FUSE became the first mission fully led by a PI and cost-capped. The Explorer program was highly active between 1995 and 2003, with six MIDEX and five SMEXs were selected for flight, though two were eventually cancelled. The program then had a drastic drop off, with no missions selected between 2003 and 2009 and a temporary cancellation.

There are currently two categories of Explorer missions: Small Explorers (SMEX), and Medium Explorers (MIDEX). SMEX missions allowed for increased levels of PI leadership while maintaining project management responsibility at GSFC (e.g. TRACE, SWAS, WIRE)[4].

SMEX and MIDEX missions had cost caps of $120 million, and $180 million, respectively [2004 numbers], adjusted for inflation at the discretion of NASA depending on the circumstances of selection timing or delays. The cost cap for each of these mission lines has been adjusted over time with each new AO. The 2008 SMEX AO limited PI Mission Cost to $105M. Note that, while launch costs are excluded from this cost, a 30% contingency is required.

Since 1988 there have been 5 SMEX AOs, and 3 MIDEX AOs. Each AO also solicits proposals for missions of opportunity (MoOs). Typically, some 30 proposed projects compete for 2 flight opportunities in each AO (see Table 3.1).

TABLE 3.1 Statistics on Proposals Submitted for Explorer Missions (from PILMSS, with 2008 details added. SOURCE: NASA, Science Mission Directorate.)

| Year | Explorer | No. of Proposals[a] | →Phase A | Selected Full Missions[c] |
|------|----------|--------------------|---------|--------------------------|
| 1997 | SMEX | 46 total, 40 full | n/a | RHESSI, **GALEX** |
| 1998 | UNEX | 29 total, 23 full | n/a | **CHIPS** (MoO), IMEX |
| 1998 | MIDEX | 31 total, 27 full | 5 | *Swift*, **FAME** (cancelled) |
| 1999 | SMEX | 33 total, 21 full | 7 | **SPIDR** (cancelled), AIM |
| 2001 | MIDEX | 31 total, 21 full | 4 | THEMIS, **WISE** |
| 2003 | SMEX | 29 total, 22 full | 5 | IBEX, **NuSTAR** |
| 2008 | SMEX | 49 total, 32 full | 6[b] | **GEMS,JANUS,TESS,**CPEX, IRIS, NICE |

*a* "Full" refers to full missions. "Total" includes mission of opportunity proposals; [b] Downselect expected Fall 2009. [c] **Astrophysics missions are in bold face.**

---

[3] http://www.planet4589.org/space/misc/explorer.html

[4] A smaller category of University-class Explorers (UNEX), with a cost cap of $15 million is on hold after only one AO, pending the availability of a suitable small (<$10 million) U.S. launch vehicle.

## Launch Rate

Many believe, as we do, that a higher astrophysics Explorer launch rate of approximately 1 per year is needed to achieve the Explorer goals of providing a rapid response to changing science and technology in order to do cutting-edge science at moderate cost.

The launch rate actually being achieved in astrophysics is 1 per 3 years. There have been 6 astrophysics launches in the 21 years since the first SMEX AO (SWAS, GALEX, WIRE, FUSE, Swift, WMAP). With a seventh due in 2009 (WISE), an eighth in 2011 (NuSTAR) and one more in astrophysics by 2015, leading to 9 launches in 27 years.

There is a long history of calls from the community for one astrophysics launch per year: Three 1991 reports had this recommendation. The Bahcall Report had the following language (p.118): *"The committee recommends that NASA increase the rate of Explorer missions for astronomy and astrophysics to six Delta-class and five SMEX missions per decade."* The Space Studies Board report says the following about the solar/heliosphere/Earth's magnetosphere Explorer program (there is no corresponding section on astrophysics): *"The recommended level of an average of one Explorer per year for solar and space physics has not been reached, however, because cost overruns in the current Explorer program continue to cause delays."* Finally, the Office of Space Science and Applications Strategic Plan had the following program guidelines: *"We endeavor to start a small mission or a small mission program every year, in conjunction with either a major or moderate mission."*

Within NASA, the earliest NASA budget request available online (1999), NASA, less explicitly, stated that *"The goal of the Explorer Program is to provide frequent, low-cost access to space...",* and more recent Roadmaps and Strategic Plans employ similar language. NASA has used the approximately annual explorers objective in its planning process. NASA's presentation to the SEUS and OS *"The Explorer Planning Budget will support the missions currently under development plus • 2 MIDEX & 2 SMEX every 3 years (planning program) ..."* (Paul Hertz, July 2, 2003). This expected SMEX/MIDEX launch rate of 2 each per 3 years, or roughly one SMEX per year and a MIDEX every other year, split evenly between Sun-Earth Connection and Astrophysics was not however realized in practice.

A major conclusion of this White Paper that the realization of the long-standing goals of both the external review committees and of NASA itself, is now timely for adoption as a major NASA goal in the decade 2010-2020.

## A. Science Gaps: Cutting-edge Science on a Budget

If Explorers fulfilled their three other roles – rapid implementation, innovation, and engaging & training personnel - but did not do front rank science, then the program should not exist. In fact, Explorers have made breakthroughs in many fields of astrophysics. The Explorer program has distinguished history, present, and future, doing cutting-edge science that simply cannot be done by facility-class instruments. The early Explorer program included UHURU, and the restructured program post-1988 included COBE, the science from both of which contributed to three of the Nobel Prizes awarded for NASA-based science – Giacconi (2002), Mather, and Smoot (2006).

### Post-1988 Science

Despite the limited launch rate since the 1988 MIDEX/SMEX re-structuring, the astronomy

Explorer missions have covered many fields of astronomy: millimeter spectroscopy (SWAS), UV and far-UV imaging and spectroscopy (FUSE, GALEX), cosmology using the cosmic microwave background (WMAP), gamma-ray bursts (Swift), with only one payload failure (WIRE).

Post-1988 Explorer accomplishments include:

***WMAP:*** marked the definitive beginning of precision cosmology:
- Age of Universe, 13.73±0.12 Gyr;
- Baryon fraction, $\Omega$(baryons) = 4.6±0.1% of the universe.
- $\Omega$(dark matter) = 23.3±1.3%;
- $\Omega$(dark energy) = 72.1±1.5%;
- Spacetime curvature is within 1% of Euclidean, improving on the precision of previous award-winning measurements by over an order of magnitude;
- Set the epoch of re-ionization to z~15

***FUSE:***
- Detected of $He^+$ re-ionization at redshifts less than;
- Traced the 'missing baryons' in the local intergalactic medium;
- Challenged models of Galactic chemical evolution. (high values of D/H in the Milky Way disk.)
- Eliminated starbursts as significant contributors to IGM ionization at the present epoch. (Very low escape fractions of Lyman continuum.)
- Found $H_2$ in nearly all sightlines through the Galactic ISM and halo, substantially increasing the mass of $H_2$ out of the Galactic plane.

***GALEX:*** Discovery of -
- Local starburst analogs of high redshift Lyman Break Galaxies;
- Stellar tidal disruption flares from otherwise inactive massive black holes;
- The remarkable turbulent wind wake of Mira and its cousin, CW Leo;
- New regimes of star formation: extended disks, primordial HI clouds, tidal tails;
- Uniformly observed star formation history of Universe over the last 7 Gyrs.

***SWAS:*** found -
- Water vapor is present in almost all star-forming regions
- Gas-phase water abundance peaks near molecular surfaces;
- Molecular clouds have an abundance of water-ice in their interiors .
- 

***Swift***, the #1 ranked mission from the 2008 Senior Review in science/dollar:
- discovery that short gamma-ray bursts are caused by compact star mergers
- first detection of an X-ray flash from the shock break out from a supernova
- metallicity measurements to redshifts z>6 using gamma-ray bursts

Explorers can also directly complement the flagship missions: wide area surveys of the sort done by GALEX, Swift/BAT and, soon, by WISE, simply cannot be done by the flagship missions. Explorer class surveys are, moreover, essential for finding the best targets so that the full potential of NASA's flagship Observatories can be realized. They serve in the same way that the 48-inch Palomar Schmidt Sky Survey did for the Palomar 200-inch, and for many later large telescopes, and as does the SDSS does today.

Earlier Explorers also had great impact:

***COBE:*** Nobel Prize winning science -
- CMB spectrum found to be a black body to high accuracy: 2.725±0.002K;
- 1:100000CMB fluctuations, consistent with CDM cosmology;
- First IR background constraints on models of the history of star formation and the buildup over time of heavy elements and dust.

***RXTE:*** Discovery of:
- the fastest oscillations known from astrophysical sources (kHz QPOs) ;
- changes on dynamical time-scales in stellar-mass black holes (ms QPOs);
- nuclear powered pulsars, (pulsations during thermonuclear X-ray bursts);
- connections between radio jet formation and accretion disk instabilities.
- frequencies that scale inversely with the mass from stellar to supermassive black holes.

***EUVE:***
- First survey of sky at 'unobservable' EUV wavelengths
- Detection of thermal EUV emission from neutron star surfaces
- First measurement of stellar coronal abundances.
- hot hydrogen-rich white dwarf atsmopheres opacity due to high $Z$ elements, not He.

## Continuing Strength

There is strong evidence that there continue to be many excellent new ideas for Explorer-class missions. This can be easily found by the 15:1 oversubscription rate of every AO, and by looking at the projects now in development: NuSTAR and WISE in phase C/D (design/build), and three, GEMS, JANUS and TESS, now in Phase A (concept study) prior to down-selection.

***WISE,*** PI: Edward Wright (UCLA, Los Angeles CA), is a MIDEX which will provide an all-sky survey from 3 to 25 μm with 500 times the sensitivity of IRAS. The 6-month survey, beginning early 2010, will help search for the origins of planets, stars, and galaxies and create an enduring legacy infrared atlas. WISE will find the most luminous galaxies in the Universe; find the closest stars to the Sun; detect most Main Belt asteroids larger than 3 km; enable a wide variety of studies ranging from the evolution of planetary debris disks to the history of star formation in normal galaxies; and will provide an important source catalog for JWST.

***NuSTAR***, PI Fiona Harrison (CalTech, Pasadena CA), is a SMEX, now scheduled for launch in mid-2011 that exploits innovative technology, graded multi-layer mirrors. Multi-layer mirror technology uses resonant Bragg reflection to create focusing optics up to energies as high as 80 keV, so reducing background by 1-2 orders of magnitude. NuSTAR will investigate key questions arising from earlier work: defining the population of supermassive black holes in galactic nuclei and the creation and dispersion of the elements in supernovae, and will explore genuinely new territory -- discovering hard X-ray sources at far fainter fluxes, and in far greater numbers than have been accessible before.

Six SMEX missions are now in Phase A studies pending a downselect in the Fall of 2009[5].

---

[5] http://www.nasa.gov/home/hqnews/2008/may/HQ_C08029_SMEX_Awards.html

Three are non-solar astrophysics missions, and so lie within the Astro2010 remit:

**Gravity and Extreme Magnetism SMEX (GEMS),** PI: Jean Swank (NASA GSFC, Greenbelt, MD) - GEMS will use an X-ray polarimetry to track the flow of highly magnetized matter into supermassive black holes.

**Joint Astrophysics Nascent Universe Satellite (JANUS),** PI: Peter Roming (Penn State University, University Park, PA) - JANUS will use a gamma-ray burst monitor to point its infrared telescope at the most distant galaxies to measure the star-formation history of the universe.

**Transiting Exoplanet Survey Satellite (TESS),** PI: George Ricker (MIT, Cambridge MA) - TESS will use a bank of six telescopes to observe the brightest 2.5 million stars and discover more than 1,000 Earth-to-Jupiter-sized planets around them.

# TECHNICAL OVERVIEW

## B. Rapid Implementation

The Explorer program can react to new scientific discoveries with much shorter turn-around than large missions (or decadal studies). Low launch rates threaten this rapid response.

*Swift* is an example. The detection of X-ray afterglows from Gamma-ray bursts (GRBs) by the Italian-Netherlands mission Beppo-SAX[6] led to the first identification of a GRB a year after launch (GRB 970402 **)**  and settled the 24-year old dispute - are GRBs isotropic because they are very nearby, or because they are cosmologically distant? – in favor of powerful events in distant galaxies. But Beppo-SAX was not designed with GRB identifications in mind. A dedicated mission could go from a handful of identified GRBs to hundreds. Within 5 years Swift had been successfully launched, and now, 5 years later, has achieved its promise, with 236 GRBs identified, 57% with redshifts, including one at z=5.6 (GRB060927), and a wealth of new GRB physics being revealed.

WMAP gives a slightly less dramatic but useful example. The extraordinary COBE results on the spectrum and fluctuations of the CMB were reported in 1990-1992. The MAP proposal was made in 1995, and selected in 1996, with launch following 5 years later, in March 2001.

## C. Innovation:

The Explorer program provides a quick path to scientific use for newly-developed technology, and thereby allow quick exploration of entirely new territory as soon as the technology is ready. This technology is often then used in other, larger programs. Recent examples programs include:

**WMAP:** was the first mission to demonstrate the viability of operating at L2. WMAP also made extensive use of passive cooling, developing and validating many techniques for some of the design concepts adopted by JWST.

**FUSE, GALEX:** there is a direct line from the development of microchannel plate detectors and holographically ruling aberration corrected gratings on sounding rockets to GALEX and FUSE, and ultimately to the Cosmic Origins Spectrograph (COS) for HST. COS would not exist without FUSE.

**NuSTAR:** Multilayers for X-ray and EUV optics were first tested on sounding rockets imaging the solar corona in the 1980's, and led to the SMEX satellite TRACE, launched exactly 10 years ago, which led in turn to the much larger set of telescopes known as AIA, about to be launched on the Solar Dynamics Observatory. The high energy mirrors of NuSTAR are closely related. NuSTAR mirrors are pathfinders for the IXO Hard X-ray Telescope.

**Swift:** Swift has pioneered the use of extremely rapid and autonomous response to celestial events and community submitted Targets of Opportunity. Swift has reacted within 52 seconds to newly discovered GRBs from the on-board detectors, and within 43 minutes to ground-based responses from other observatories. Swift helps all missions by demonstrating a 'lights-out monitoring' capability to keep operations cost at a minimum, while still responding to astronomical discoveries on a 24/7 basis.

---

[6]http://heasarc.gsfc.nasa.gov/docs/sax/sax.html

## D. Engage and train*:*

The NASA Explorer program has wider value, beyond the immediate science return, to both the space program, and to the technological health of the nation. By first engaging the imagination of scientists and engineers, and then providing broad, hands-on, training, the Explorer program adds human capital to NASA and the nation.

The PILMSS report recognized this: *"The science return [of the Explorer program] is in fact much broader than just the new knowledge about space science enabled by a specific mission, and the PI-led programs play a particularly important role in these crucial, ancillary aspects:*
*(1) **training the next generation** of scientists and engineers,*
*(2) **strengthening the scientific and technical infrastructure**, including instrument and spacecraft developers, launch services, and the institutions that manage this complex enterprise,*
*(3) **generating excitement in the science and larger communities** via the PI-led project team's enthusiastic promotion of the mission."* (p.47, emphasis added.)

The Explorer program itself supplies numerous training stories. E.g. Charles Bennett was Deputy PI for DMR on service on COBE, which gave him the experience necessary to be PI of WMAP. Similarly, Ned Wright was also a COBE co-I and is now PI for WISE.

The Explorer program also enlarges the pool of commercial companies capable of building spacecraft by involving more institutions and commercial companies in the scientific space program. Engineers and management at the various industrial partners are always enthusiastic about Explorers, even though they freely acknowledge that they will make little money from these projects. These missions engage and challenge the engineers in ways that their commercial business generally doesn't.

The relatively low current Explorer launch rate limits these benefits. Riccardo Giacconi has recently stated[7] that: *"NASA has not achieved the balance between very large, medium and small missions that the community has advocated for years. This lack of smaller, principal investigators and university led missions makes it very difficult to train the experimental scientists of tomorrow."* It is significant, then, that a number of industrial partners, such as General Dynamics, decided not to participate in the most recent (2008) SMEX round because the cost cap was too tight.

A vigorous Explorer program, with much more frequent launches than over the past two decades, will accelerate all of these important training and engagement benefits, growing the aerospace science, management and engineering capability available to NASA, and to the nation.

---

[7] Astro2010 State of the Profession White paper, Giacconi_management_FFP_IPP[1].pdf

# TECHNOLOGY DRIVERS

## 1. Launcher and Spacecraft Costs and Capabilities:

### *Launcher:*

The DOD has recently moved from the Delta II rocket to the significantly more capable, but also more expensive, Delta IV and Atlas V rockets for military launches. NASA cannot afford to support the Delta II program on its own, and now has only light rockets (Pegasus, 450 kg to LEO, and Taurus[8], 1350 kg to LEO) and heavy lift vehicles (Delta IV, Atlas V, 8.6-29.4 tonnes to LEO) as options for launches. Explorers are sensitive to this gap in lift capability because, after subtracting the mass of the spacecraft, there is little mass remaining for instruments on light vehicles, while a single heavy lift vehicle is as expensive as the entire Explorer budget. We encourage NASA to seek new launch capabilities to fill this strategic gap in moderate lift rockets to enable the Explorer program to continue launching missions with significant science payoff but reduced costs and timescales compared to large strategic missions. We are encouraged by the planned use of a Minotaur IV by NASA for the upcoming LADEE lunar mission.

For light launchers, the high cost of Pegasus and Taurus puts a heavy burden on Explorers. While NASA is strictly prohibited from buying foreign launchers, their pricing structure sets a plausible goal, with Taurus-class ~1-tonne payloads to LEO available in the $10M-$18M range[9]. There are cheaper, and somewhat more capable, US built launch vehicles available either now or in the near-term[10]: the Minotaur (based on the Minuteman ICBM with Taurus upper stages,), and the new SpaceX Falcon 1. The Minotaur family can put 580kg (Minotaur I) to 1,700 kg (Minotaur IV, equivalent to a Delta-II) into LEO. Seven USAF Minotaur launches have taken place so far, all successful. The liquid-fueled Falcon 1 has achieved LEO and is currently anticipated to deliver 420kg (Falcon 1) to 1010 kg (Falcon 1e) to LEO for $7.9M to $9.1M.

### *Spacecraft:*

The costs for a basic spacecraft bus, integration and test, a control center, ground station access, data processing, and a minimal science team, leaves little money left for an instrument payload. The mission assurance requirements imposed on the spacecraft have a large impact on the spacecraft cost. Yet, the S/C bus and some of the support electronics and software are off-the-shelf items, most subsystems being highly standardized. Is there scope for reducing costs here? Relaxation of NASA EEE parts assurance requirements to levels acceptable by commercial customers may result in substantial savings, including indirect ones, as e.g. increased radiation-hardness costs in mass and power. While taking shortcuts in performance or qualification testing adds risk, but needs to be balanced against the larger risk that comes from the most innovative subsystems of the mission, which tend to be the ones in the payload. We urge NASA to welcome and encourage initiatives to reduce spacecraft costs e.g. the CREST Astro2010 White Paper[11].

---

[8] http://www.orbital.com/SpaceLaunch/Pegasus/, http://www.orbital.com/SpaceLaunch/Taurus/

[9] PSLV: www.antrix.gov.in; Rokot: www.eurockot.com; Kosmos-3M: www.cosmos-space.de; Dnepr: www.kosmotras.ru; Vega: http://www.arianespace.com/launch-services-vega_overview.asp

[10] http://www.orbital.com/SpaceLaunch/Minotaur/, http://www.spacex.com/falcon1.php

[11] *"Center for Research on Experimental Satellite Technology"*, Chakrabarty et al. (2009).

## 2. Management and costing:

The Explorer program has the potential to provide innovation in management processes. The PILMSS wrote: "***Finding.*** *The space science community believes that the scientific effectiveness of PI-led missions is largely due to the direct involvement of PIs in shaping the decisions and the mission approach to realizing the proposed science concepts.*" The correct PI-NASA balance is hard to find. NASA naturally tends towards taking major roles in the Explorers to minimize risk.

### *Analyzing Risk:*

To allow for the risks and resulting cost growth in Explorers, NASA requires a 30% Contingency Funds for Explorers. In small cost-capped missions, with fixed launch and inflexible spacecraft costs, a blanket 30% contingency has a disproportionate impact on the scientific payload. On the surface, 30% is a reasonable figure, based on "The average increase in development costs for PI-led Explorers is 30.8 percent compared with the 18.3 percent for the other recent missions, though just two cases—Swift and GALEX—are the cause of the high average." (PILMSS)

Another approach would be to diagnose the most common causes of overruns and then try to minimize them. If validated, the Explorer contingency could then be prudently reduced. Much of the current payload risk stems from technology development challenges. The PILMSS "*examined five Explorer missions (RHESSI, Swift, IMAGE, GALEX, and WMAP) (see their* Table 5.2*); on average, their development cost increased 30.8 percent. Two of the missions, Swift and GALEX, showed particularly large increases in development costs, 68.8 percent and 52.8 percent, respectively, over their selection cost caps. Both Swift and GALEX encountered problems related to immature technology, which probably contributed to the cost and schedule overruns ...*". The PILMSS report (Chapter 5, p.39) concluded that NASA could address these technology development overruns by taking action in two areas: pre-proposal, and phase-A studies. The 2008 Phase A studies were just $750k. We re-iterate their recommendations (emphasis added):

**PILMSS recommendation #5**. "*NASA should set aside meaningful levels of regular funding in PI-led programs to sponsor relevant, competed **technology development** efforts. The results from these program-oriented activities should be made openly available on the program library Web site and in articles published in journals or on the World Wide Web.*"
**PILMSS recommendation #2.** "*NASA should **increase the funding for and duration of concept studies** (Phase A) to ensure that more accurate information on cost, schedule, and technical readiness is available for final selection of PI-led missions.*"

We agree with this assessment, and further link adequate technology development for Explorers to the existence of a healthy sub-orbital program. We urge NASA to thoroughly review the Explorer development and contingency rules with the goal of fulfilling these recommendations.

### Reporting and Review Requirements

The same changes that led to increased cost contingencies for Explorers also led to the imposition of extra reporting and review requirements. While these are appropriate to large programs, applying identical standards to the Explorer programs has imposed a significant burden, that is seen as significant by the small PI teams, and possibly as counter-productive, as they re-direct resources from solving project issues.

Again, the PILMSS has stated the case clearly: *"Particularly in the late 1990s and early 2000s, NASA made many changes to its oversight philosophy and its views on the acceptability of risk in response to a series of failures and mishaps (not, however, in PI-led missions). ... **new agency-wide requirements** that affected all NASA missions, **including PI-led missions**. The major changes that NASA mandated and that were incorporated into subsequent AOs include the addition of software IV&V and **new risk management and cost reserve requirements**. ... In addition... NASA... added oversight in the form of formal reviews beyond those agreed upon after selection... . ... projects reported that they were increasingly required to present formal reviews on various technical or management issues raised by NASA officials... the burden of preparing, holding, and following up on action items—was **significant in the view of many PIs and PMs. They said these formal reviews (as opposed to peer reviews) had minimal value and were burdensome because they took time away from critical project activities."* [PILMSS, ch.4, p.37: "Changes in Management and Oversight"**, emphasis added**.]

We concur with the PILMSS recommendation #9:

**PILMSS recommendation #9:** *NASA should resist increasing PI-led mission technical and oversight requirements—as, for example, on quality assurance, documentation, ITA-imposed requirements, or the use of independent reviews—to the level of requirements for larger core missions and should select missions whose risks are well understood and that have plans for adequate and effective testing.*

Prudent lessening the oversight requirements on Explorers from those appropriate for large mission should be part of any review of management practices for Explorers.

## Phase B to Launch

Explorers have averaged 4-6 years from start of Phase B to launch. It is important to maintain this tight schedule. Three years from the start of Phase B to launch is a very tight schedule, with little of the schedule margin TMCO reviews look for, and which PI teams all like to have. On this schedule, contracts for long-lead items must be ready to sign on day 1. A more thorough Phase A will make it more likely that the PI will be able to keep to such a schedule. Conversely, 4 years should be sufficient time for a SMEX, and 5 years for a MIDEX. On the other hand, prolonging a development program is a sure way to run past a cost cap. This is especially true when it is due to the schedule being extended.

## Program Integrity and Continuity

Both WISE and NuSTAR were nearly eliminated in 2006, in order to fund shortfalls in other programs. NuSTAR, the most recently selected SMEX, was abruptly cancelled in Feb 2006, two weeks before its initial confirmation review, and the 2006 budget of WISE, the most recently selected MIDEX, was cut in March 2006 by over 50%. Neither WISE nor NuSTAR were experiencing internal budget or technical problems. Eventually, WISE survived and NuSTAR was restarted in Sept 2007. Although WISE benefited from its extended Phase A study, the major budget cutbacks in 2006 led to a 17-month launch delay, and substantial additional cost.

We urge NASA to strive hard to protect the funding for missions that have reached Phase B, to prevent significant launch slips and cancellations, with a goal of 4 to 5 years from Phase B to launch.

# ORGANIZATION, PARTNERSHIPS & CURRENT STATUS

We believe that Explorers, as the largest PI-led programs in astrophysics at NASA, form a crucial part of a continuum of innovation, from lab development and sub-orbital programs on smaller scales, through to facility-class programs on larger scales. Without the component parts being healthy, the program as a whole will eventually falter and the nation's capability to build epoch-making large missions will wither.

Explorers are the largest of the PI-led endeavors within NASA. Most technologies used in Explorers, and later used in large missions, develop out of initially small programs based on the ideas and initiative of a single PI. The Technical Readiness Level (TRL) formalism[12] helps clarify this development path. From initial idea (TRL-1, 'basic principles observed') to TRL-6 ('system/subsystem demonstration in space') to the final TRL-9 ('flight proven') maps to:

- ***Tech. development:*** taking TRL-1 to TRL-4 ('lab. Demonstration at component level') or TRL-5 ('lab. Demo in appropriate environment')
- ***Sub-orbital balloon, rocket:*** reaches TRL-5/-6
- ***Explorer:*** TakeTRL-5/-6 to TRL-7 in Phase A, TRL-9 for flight.

The status of Explorer development in this framework has limitations, as described in the previous sections. Funding for basic technology development is limited (details), while the sub-orbital program is at a historical low point. Other White Papers address these areas.

The goal of one astrophysics Explorer launch per year matches well with the, independently derived, goal of the ASRAT Rocket Program White Paper. They say: *"So what does the community require? GALEX provides a case study. They employed 6 Ph.D. instrumentalist and required the development of ~ 4 new technologies with sounding rocket heritage. So the combined yearly output of 12 sounding rocket programs, assuming **no attrition,** is just able to sustain the workforce and technical needs of 1 Explorer program per year."*

A strength of the Explorer program is the wide range of institutions involved. The list of institutions of the co-authors of this White Paper illustrates this.

---

[12] http://www.hq.nasa.gov/office/codeq/trl/

# COST ESTIMATES

## *NASA Explorer Budget*

The Explorer program was highly active between 1995 and 2003. During those years six MIDEX and five SMEXs were selected for flight though two were eventually cancelled.

The approximate total costs, including program management overhead, was ~$1700M. The average program expenditure during these years was about $200M real year dollars per year.

The program then had a drastic drop off with no missions selected between 2003 and 2009. The result was a decrease in budget to less than ~$10M per year. There is new life in the program with NuSTAR moved to flight status and SMEX proposals currently under review.

The budget projections are currently in the $150M per year range for the next five years.

## *SMEX Budget: a worked example*

Below we give a sample breakdown for major categories from a recent SMEX proposal. The individual numbers do not match the actual proposal, but rather are average numbers from several sources for each category. The details of the payload were unique to this mission, of course, but there was no new technology, or any particular performance requirement that had not been achieved in other modest missions. The total is $10M over the cost cap. There was no guest observer program, and data archiving and distribution were provided by MAST with only a very small cost charged to the mission.

There are many ways to break down these costs, of course. A somewhat more capable than average spacecraft was required; a more basic S/C would have come in around $32M. Fairly consistent quotes for the S/C came from three different vendors. Other contributions to the costs were roughly independent of mission characteristics.

| | | |
|---|---|---|
| Phase A | $0.75M | |
| Project management | $3.5M | |
| Systems engineering | $1.5M | |
| Mission Assurance | $1M | |
| Spacecraft bus | $36M (incl. S/C engineering and S/C I&T) | |
| Payload | $28M | |
| (incl. H/W, engineering, instr. I&T) 40-cm telescope, standard detectors, some custom electronics | | |
| Satellite I&T | $2.5M | |
| Flight Ops Center | $7.5M | |
| includes: pre-launch development | $2.5M | |
| flight operations (2yrs) | $3.5M | |
| Ground station (antenna usage) | $1.5M | |
| Science Data processing[a] | $3M | |
| Science team | $5M | |
| (incl. E/PO) | | |
| Launch operations | $1M | |
| **Total:** | **$89.5** | |
| 30% reserves (not applied to science team): $25.4 | | |
| **Grand total:** | **$114.95** | |

a. Includes: data processing pipeline development. post-launch processing, archiving, distribution, data calibration.

# ACTIVITY SCHEDULE

*"NASA is an evolving organization that learns from past experience, events, and new information and responds to administrative interests and factors"*
[PILMSS, ch.4, p.37.]

We consider the, peer reviewed, PI led, Explorer, sub-orbital, and technology development programs to be all vital components of a coherent NASA development strategy to enable the long-term health of the nation's space program, including the health of large mission programs.

We recommend and urge that NASA, in order to create a vigorous Explorer program:

**1. Achieve the Long-advocated Astrophysics Explorer Flight Rate of 1/year.**
to provide the low-cost, rapid response, new technology and training resource that the Explorer program was always intended to be. We urge NASA to include an enhanced Explorer program in its FY11 budget request.

**2. Create a Task Force on 'Innovative Approaches to Small Missions'.**
This task force should involve NASA, PI-class scientists, spacecraft vendors, launch companies. The Task Force should consider at least:
    A.  Can reporting/review procedures be reduced, simplified?
    B.  Can contingency reserves be lowered by funding technology development and longer, stronger Phase A studies.
    C.  Should contingency reserves apply equally to all mission elements?
    D.  How can NASA encourage savings using external groups?
    E.  A stronger sub-orbital program to raise the TRL of pivotal technologies, so mitigating the risk of using such technologies in Explorer missions.
    F.  Is SMEX/MIDEX distinction still useful? Is the SMEX program is meeting its intended objectives?
    G.  Would Wallops (WFF) launches enable streamlined procedures?
    H.  Reassess the SMEX cost cap based on what was learned from items A-G.
The Task Force should complete its report within a year, to allow NASA to incorporate its findings into FY2011 planning.

**3. Take an active role in fostering capable, low cost, US launch vehicles.**
We envisage medium-class LV, in the $10M-$18M cost bracket, capable of delivering 1000 kg to LEO, and possibly even to Earth trailaway or L2 orbits. Non-US launchers of this capability exist, and challenge the US dominance of the space industry.

**4. Make a strong investment in the personnel side of the Explorer program.**
The people and teams trained as part of the Explorer program have the real world experience to take them into the next league, and produces the new project leaders and PIs who will be needed when the current baby-boomer generation retires. There simply aren't enough new people coming up through the system to take their place. Without this training NASA will be left with "green", unseasoned, people running large projects

**6. Follow the PILMSS recommendations** [Appendix A].

## Appendix A: PILMSS recommendations:

1. NASA should consider modifying the PI-led mission selection process in the following ways:
* Revise the required content of the mission proposals to allow informed selection while minimizing the burden on the proposing and reviewing communities by, for example, reconsidering the TMC-lite approach and eliminating the need for content that restates program requirements or provides detailed descriptions such as schedules that would be better left for post-selection concept studies,
* Alter the order of the review process by removing low- to medium-ranking science proposals from the competition before the TMC review, and
* Allow review panels to further query proposers of the most promising subset of concepts for clarification, as necessary.

2. NASA should increase the funding for and duration of concept studies (Phase A) to ensure that more accurate information on cost, schedule, and technical readiness is available for final selection of PI-led missions.

3. NASA should make explicit all factors to be considered in the selection of PI-led missions—for example, targets and/or technologies that are especially timely and any factors related to allocating work among institutions and NASA centers.

4. NASA should develop PI/PM teams whose combined experience and personal commitment to the proposed implementation plan can be evaluated. NASA should also provide opportunities for scientists and engineers to gain practical spaceflight experience before they become involved in PI-led or core NASA missions. These opportunities could become available as a result of revitalizing some smaller flight programs, such as the sounding rocket and University-class Explorer programs.

5. NASA should set aside meaningful levels of regular funding in PI-led programs to sponsor relevant, competed technology development efforts. The results from these program-oriented activities should be made openly available on the program library Web site and in articles published in journals or on the World Wide Web.

6. NASA and individual mission PIs should mutually agree on a funding profile that will support mission development and execution as efficiently as possible. If NASA must later deviate from that profile, the mission cost cap should be adjusted upward to cover the cost of the inefficiency that results from the change in funding profile (see Recommendation 10).

7. NASA PI-led-mission program officials should use recent experiences with ITAR to clarify for proposers (in the AO) and for selected projects (e.g., in guidance on writing technical assistance agreements and transferal letters) the appropriate application of ITAR rules and regulations.

8. NASA should ensure stability at its program offices, while providing sufficient personnel and authority to enable their effectiveness, both in supporting their missions and in reporting to and planning with NASA Headquarters.

9. NASA should resist increasing PI-led mission technical and oversight requirements—as, for example, on quality assurance, documentation, ITA-imposed requirements, or the use of independent reviews—to the level of requirements for larger core missions and should select missions whose risks are well understood and that have plans for adequate and effective testing.

10. NASA should clarify the change-of-scope procedures available for projects to negotiate the cost and schedule impacts of any changes in requirements initiated by NASA Headquarters.

11. NASA should continue to use the existing termination review process to decide the fate of PI-led missions that exceed their cost cap. It should develop lessons learned from termination reviews and make them available to other PI-led projects.

12. NASA should not descope mission capabilities (including science instruments) without the PI's agreement or outside the termination review process.

# Appendix B: Explorer Missions 1989-present (From PILMSS)

| No. | Name | Class[a] | Principal Investigator[b] | Notes[c] |
|---|---|---|---|---|
| 66 | Cosmic Background Explorer (COBE) | | Early-style PI mission-science lead: John Mather, GSFC | CMBR, anisotropy, infrared |
| 67 | Extreme Ultraviolet Explorer (EUVE) | | Early-style PI mission-science lead: Stuart Bowyer, UC Berkeley | EUV full-sky survey, deep survey, ISM |
| 68 | Solar Anomalous and Magnetospheric Particle Explorer (SAMPEX) | SMEX | Early-style PI mission-science lead: Glenn Mason, University of Maryland | Cosmic rays, magnetosphere |
| 69 | Rossi X-Ray Timing Explorer (RXTE) | Explorer | Early-style PI mission-science lead: Richard Rothschild (UCSD), Jean Swank (GSFC), Hale Bradt (MIT). | X-ray spectral phenomena of stellar and galactic systems |
| 70 | Fast Auroral Snapshot Explorer (FAST) | SMEX | Early-style PI mission-science lead: Charles Carlson, UC Berkeley | Plasma physics, aurora |
| 71 | Advanced Composition Explorer (ACE) | Explorer | Early-style PI mission-science lead: Ed Stone, Caltech | Solar corona, IPM, ISM, cosmic rays |
| 72 | Student Nitric Oxide Explorer (SNOE) | UNEX/STEDI | Charles Barth, University of Colorado | Nitric oxide density fluctuations in thermosphere due to solar and auroral activity |
| 73 | Transition Region and Coronal Explorer (TRACE) | SMEX | Early-style PI mission-science lead: Alan Title, Lockheed Martin | Solar photosphere, magnetism, flares |
| 74 | Submillimeter Wave Astronomy Satellite (SWAS) | SMEX | Early-style PI mission-science lead: Gary Melnick, HSCA | Interstellar clouds, star/planet formation |
| 75 | Wide-Field Infrared Explorer Mission (WIRE) | SMEX | Early-style PI mission-science lead: Perry Hacking, JPL | Protogalaxies at different redshifts (technical failure) |
| 76 | Tomographic Experiment Using Radiative Recombinative Ionospheric EUV and Radio Sources (TERRIERS) | UNEX/STEDI | Daniel Cotton, Boston University | Ionosphere electron density/photo-emission (technical failure) |
| 77 | Far Ultraviolet Spectroscopic Explorer (FUSE) | Explorer | Warren Moos, Johns Hopkins University (PI-led after design definition phase) | Origin/evolution of deuterium, galaxies, stars in far UV |

TABLE 2.1 Explorer Missions, 1989-Present